\UseRawInputEncoding
\documentclass[prresearch,twocolumn,superscriptaddress,floatfix,nopacs,nofootinbib,longbibliography]{revtex4-2}

\usepackage{graphicx,amsfonts,amssymb,amsmath,hyperref,enumerate}

\usepackage{pgfplots}
\usepackage{pgf}
\usepackage{lmodern}
\usepackage{import}
\usepackage{xr}
\usepackage{enumitem}
\usepackage{soul}
\usepackage[normalem]{ulem}
\usepackage{multirow}
\usepackage{xcolor}

\newif\ifhyper
\hypertrue
\ifhyper
\hypersetup{
   citecolor = {red},
   colorlinks = {true}, 
   urlcolor = {blue} 
}
\fi

\newcommand{\beq}{\begin{equation}}
\newcommand{\eeq}{\end{equation}}
\newcommand{\beqa}{\begin{eqnarray}}
\newcommand{\eeqa}{\end{eqnarray}}

\newcommand{\comment}[1]{}

\definecolor{addgreen}{rgb}{0.0,0.45,0.0}

\def\Longarrow{\protect\@lra}
\def\@lra{\relbar\joinrel\relbar\joinrel\relbar\joinrel\relbar\joinrel\rightarrow}

\begin{document} 

\title{Quantum Large Language Models via Tensor Network Disentanglers}

\author{Borja Aizpurua}

\affiliation{Multiverse Computing, Parque Cientifico y Tecnol\'{o}gico de Gipuzkoa, Paseo de Miram\'{o}n, 170 3$^{\,\circ}$ Planta, 20014 Donostia / San Sebasti\'{a}n, Spain}
\affiliation{Department of Basic Sciences, Tecnun - University of Navarra, E-20018 San Sebasti\'an, Spain}

\author{Fernando Lor\'{e}n}
\affiliation{Multiverse Computing, Avenida de G\'{o}mez Laguna 25, 7B, 50009 Zaragoza, Spain}

\author{Saeed S. Jahromi}
\affiliation{Multiverse Computing, Parque Cientifico y Tecnol\'{o}gico de Gipuzkoa, Paseo de Miram\'{o}n, 170 3$^{\,\circ}$ Planta, 20014 Donostia / San Sebasti\'{a}n, Spain}
\affiliation{Donostia International Physics Center, Paseo Manuel de Lardizabal 4, E-20018 San Sebasti\'an, Spain}

\author{Sukhbinder Singh}
\affiliation{Multiverse Computing, Centre for Social Innovation, 192 Spadina Avenue Suite 509, Toronto, ON M5T 2C2 Canada}

\author{Rom\'{a}n Or\'{u}s}
\email{roman.orus@multiversecomputing.com}

\affiliation{Multiverse Computing, Parque Cientifico y Tecnol\'{o}gico de Gipuzkoa, Paseo de Miram\'{o}n, 170 3$^{\,\circ}$ Planta, 20014 Donostia / San Sebasti\'{a}n, Spain}

\affiliation{Donostia International Physics Center, Paseo Manuel de Lardizabal 4, E-20018 San Sebasti\'an, Spain}

\affiliation{Ikerbasque Foundation for Science, Maria Diaz de Haro 3, E-48013 Bilbao, Spain}

\begin{abstract}
We introduce a framework for seamlessly integrating quantum computing into pretrained large language models (LLMs). The key idea is to construct a hybrid quantum-classical representation that exactly reproduces the original model, providing a principled starting point from which quantum resources can only improve performance. Our approach replaces the weight matrices in self-attention and multilayer perceptron layers with two variational quantum circuits coupled to a matrix product operator (MPO). Tensor network disentanglers transfer much of each layer's information into the quantum circuits, enabling the remaining tensor network to be compressed to a bond-dimension-one MPO with over three orders of magnitude fewer classical parameters (in our experiments, from 110,592 to approximately 36 for the replaced layer) and less than a 0.3\% increase in perplexity. Training an added unitary adapter on top of this representation then surpasses the original model, reducing perplexity by up to 1.6\%. Finally, we validate the hybrid architecture on a real quantum processor, demonstrating a practical route towards quantum-enhanced language models.
\end{abstract}


\maketitle

\section{Introduction}
\label{sec:intro}
Large language models (LLMs)~\cite{llm}, built on the transformer architecture~\cite{attention}, have rapidly become central tools of modern artificial intelligence since the release of ChatGPT~\cite{chatgpt} in 2022, and a growing number of them are openly available, including Meta's LLaMA~\cite{LlaMA} and Google's BERT~\cite{bert}. These capabilities come at a steep and rapidly growing cost: state-of-the-art models contain up to hundreds of billions of parameters, and both their training and their inference demand enormous computational and energy resources~\cite{economist}. Reducing this footprint without sacrificing accuracy has therefore become a central problem, motivating an active line of research on model compression.

A particularly effective route to compression represents the large weight matrices of the network as quantum-inspired tensor networks (TNs)~\cite{tn1, tn2}. Substituting weight matrices by matrix product operators (MPOs) has been shown to compress neural networks~\cite{tnn0, tnn1, tnn2, tnn3} and, more recently, LLMs by over $90\%$ while preserving their accuracy~\cite{compactifai}. Such representations, however, remain entirely classical: their expressivity is bounded by that of the underlying tensor-network ansatz, and once the bond dimension is fixed they cannot capture correlations beyond it.

In parallel, quantum computing has matured into the noisy intermediate-scale quantum (NISQ)~\cite{nisq} era, in which processors of hundreds of noisy qubits already run non-trivial variational quantum circuits for tasks such as quantum optimization~\cite{vqe} and quantum machine learning~\cite{qml1, qml2}. Although such processors are not error-corrected, and are therefore restricted to circuits of moderate depth, this is precisely the regime in which variational circuits are useful: their parameters are optimized in the presence of the device's own imperfections rather than assuming ideal gates. Together with advances in error mitigation~\cite{ibm} and the hardware roadmaps of major providers~\cite{ibm_quantum_roadmap_2024}, this suggests that reliable variational circuits operating on hundreds of qubits will be feasible in the near future. A natural way past the classical ceiling described above is therefore to bring the two fields together, introducing variational quantum circuits directly into the deep layers of an LLM so as to augment the expressivity of the classical (tensorized) representation beyond its fixed-ansatz limit. Early quantum approaches have focused on the simple language models~\cite{qlang1, qlang2}, however it is not yet clear how to introduce quantum resources into a pretrained LLM without degrading the model accuracy.

In this work we answer this question by encoding a pretrained LLM layer into a hybrid quantum-classical form that exactly reproduces the original model and can only be improved upon by adding and training quantum resources. The starting point is a hybrid layer in which each weight matrix is replaced by two variational quantum circuits sandwiching an MPO (VQC+TN+VQC). The central ingredient that makes this construction faithful is a pair of tensor-network disentanglers~\cite{disentanglers, ER, MERA, fmps-tjwy}, borrowed from entanglement-renormalization techniques and here applied to the MPO of a weight layer: they transfer most of the layer's information into the two quantum circuits while reproducing the original layer to numerical precision. Because the circuits are initialized so as to leave the layer unchanged, the hybrid model departs from the classical baseline rather than from a randomly initialized ansatz, so that any subsequently added quantum resources can only enhance it. We stress that the resulting quantum LLM retains the classical transformer architecture: only the weight-carrying layers are rewritten, while the surrounding network is left untouched. The construction is therefore a drop-in replacement acting on one layer at a time, and does not require the model to be retrained from scratch.

Once the layer has been disentangled, its correlations are concentrated in the two circuits and the residual tensor network becomes highly compressible. We show that it can be truncated all the way to a bond-dimension-one MPO, shrinking the classical parameter count of the layer by more than three orders of magnitude, from $110{,}592$ to $\sim36$ parameters in our experiments, at a perplexity cost below $0.3\%$. Going beyond mere re-expression of the original layer, we then insert an additional trainable unitary adapter and show that fine-tuning it surpasses the original model, lowering the perplexity by up to $\sim1.6\%$. Finally, we validate the full hybrid architecture on a compressed SmolLM2~\cite{smollm2} model and execute part of its circuits on a real superconducting quantum processor, demonstrating a practical route towards quantum-enhanced LLMs; a systematic study of the hardware behavior is reported separately in Ref.~\cite{aizpurua2026quantumenhanced}.

The remainder of the paper is organized as follows. Section~\ref{sec:method} presents the hybrid quantum-classical architecture together with the method for encoding a pretrained layer through tensor-network disentanglers, enhancing it beyond the classical baseline, and deploying it on quantum hardware. Section~\ref{sec:results} presents our numerical and hardware results on a compressed SmolLM2 model, and Section~\ref{sec:discussion} discusses the implications and concludes. Technical details of the disentangling algorithm are collected in Appendix~\ref{app:disentangling}.

\section{Model and method}
\label{sec:method}

\subsection{Hybrid quantum-classical architecture}
\label{sec:architecture}
In a previous work, we demonstrated that substituting weight matrices with TNs in LLMs can achieve over 90\% memory compression while preserving model accuracy~\cite{compactifai}, consistent with earlier findings in other AI models~\cite{tnn0, tnn1, tnn2, tnn3}.

In this paper, we build upon those results and take a further step: what if we replace the weight matrix with a variational quantum circuit? Such a replacement would enable the model to capture a significantly higher degree of correlations, far beyond what the weight matrix alone can represent. However, the unitarity of the quantum circuit would introduce constraints in the model's optimization, potentially lowering its performance. To address this limitation we propose to combine the quantum circuits with a non-unitary TN.

Thus, the weight matrices in the deep layers are replaced by (i) a Variational Quantum Circuit (VQC), followed by (ii) an arbitrary TN, such as a Matrix Product Operator (MPO), and then followed again by (iii) a second VQC. If we omit (i) and (iii), we recover the tensorized and compressed LLM models from Ref.~\cite{compactifai}.

By incorporating quantum circuits, we obtain a hybrid quantum LLM architecture that captures quantum correlations, in addition to those already present in the classical (potentially tensorized) LLM. Its baseline is therefore equal to that of the classical model and, if extra gates are added and trained, could outperform it. The associated memory overhead scales proportionally with the depth of the quantum circuit, which in the worst case grows polynomially with the number of input bits to the layer.

\begin{figure}[tb]
    \centering
    \includegraphics[width=\linewidth]{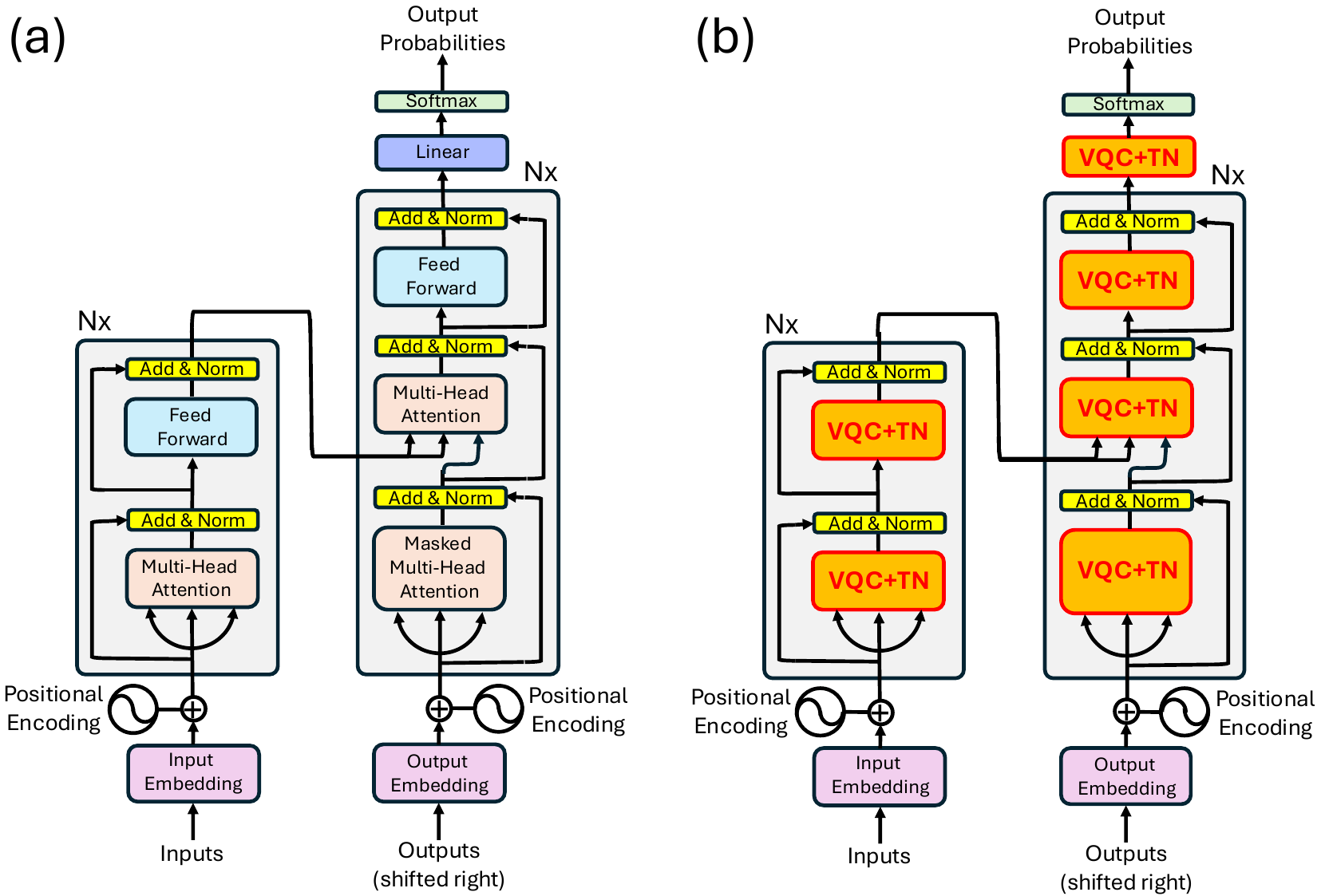}
    \caption{(a) Transformer architecture for an LLM, as discussed originally in Ref.~\cite{attention}; (b) our quantum LLM architecture.}
    \label{fig1}
\end{figure}

In Fig. \ref{fig1}, we illustrate how to practically implement this approach within the transformer architecture used in LLMs such as LLaMA. In this architecture, Self-Attention (SA) and Multilayer Perceptron (MLP) layers contain large weight matrices, particularly suitable to be replaced with the combination of VQC+TN+VQC. This procedure can also be extended to models with more complex architectures, such as mixture of experts like Mixtral8x7B~\cite{mixtral}, among others.

\subsection{Encoding a pretrained layer}
\label{sec:encoding}
To apply the concept above in practice, we could begin by training the ``empty" VQC+TN+VQC structure from scratch. However, a more efficient strategy involves leveraging existing classical LLMs (such as ChatGPT, LLaMA, BERT and Mistral) and converting them directly. Therefore, we outline a procedure to encode a classical LLM into this format and enhance its performance beyond that of the original model.

First, consider a deep layer in an LLM characterized by a weight matrix $W$. To encode it in the VQC+TN+VQC framework we follow three steps, illustrated in Fig. \ref{fig2}:

\begin{figure}[tb]
    \centering
    \includegraphics[width=\linewidth]{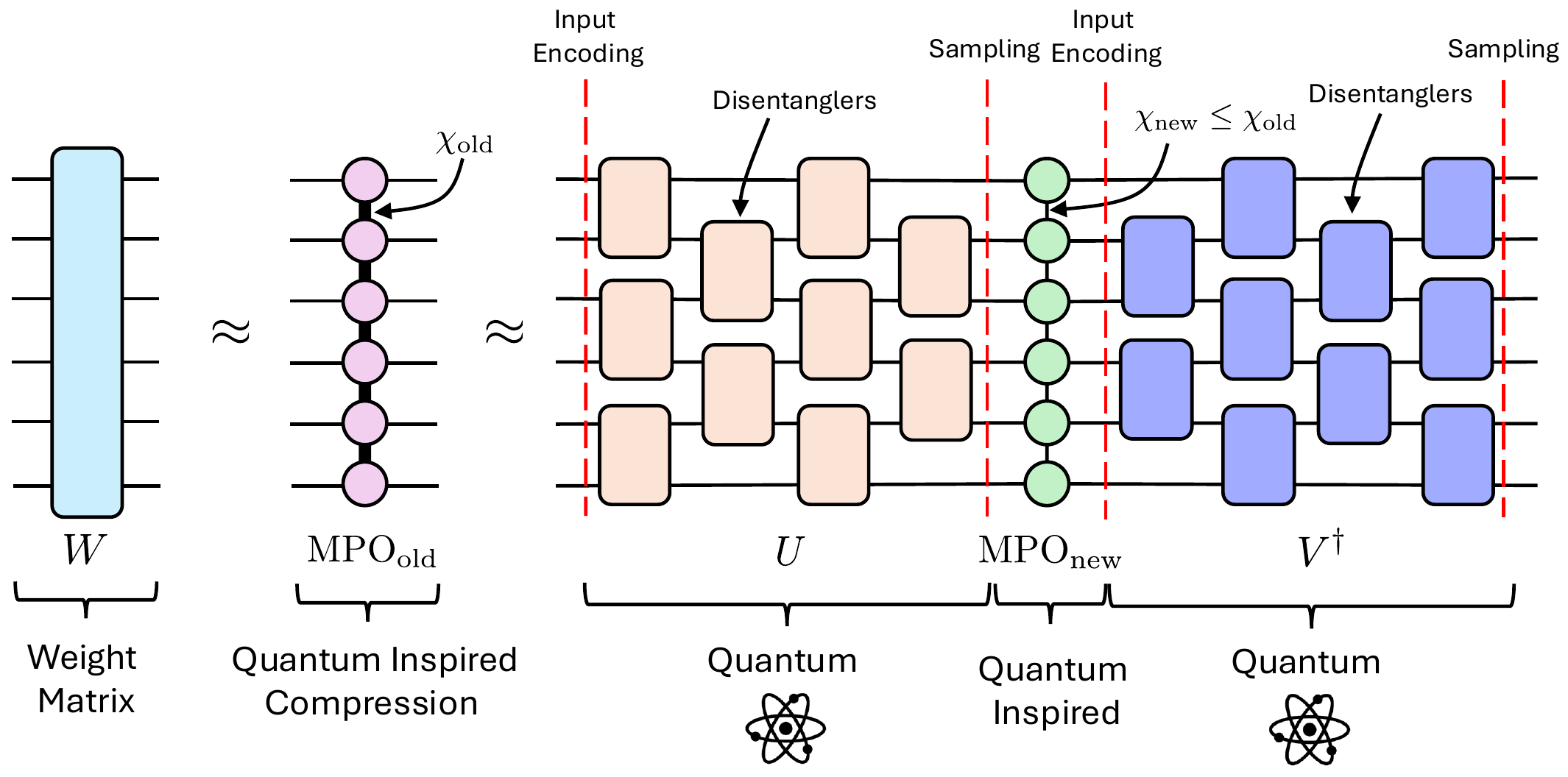}
    \caption{The weight matrix $W$ is decomposed as an MPO, which is then disentangled by the quantum circuits $U$ and $V^\dagger$, leaving a residual operator again written in MPO format.}
    \label{fig2}
\end{figure}

\begin{enumerate}
\item{This step involves implementing the TN compression algorithm from Ref.~\cite{compactifai}, which rewrites $W$ as an MPO decomposition that explicitly reveals the relevant correlations between the degrees of freedom in the layer. As a result, the memory usage of that layer is reduced due to the truncation of the bond dimension $\chi$, which effectively discards irrelevant parameters.}
\item{Compute \emph{the VQCs of disentanglers\footnote{In fact, the most general disentangler for an operator is built using generic superoperators, which may be decomposable or not in terms of Kraus operators. Quantum circuits are a particular case.}} for the input and output of the MPO. Specifically, compute two circuits composed of few-body unitary gates that remove as much entanglement as possible from the MPO. The computation of disentanglers is a well-established procedure in the field of tensor networks~\cite{disentanglers}, forming the core of techniques such as Entanglement Renormalization (ER)~\cite{ER} and the Multiscale Entanglement Renormalization Ansatz (MERA)~\cite{MERA}. We follow the explicit variational disentangling algorithm of Ref.~\cite{fmps-tjwy}, originally developed for classical image-classification models, which we summarize together with the circuit-ansatz choices in Appendix~\ref{app:disentangling}. Concretely, each gate is updated from its environment tensor through a singular value decomposition, and the two circuits are swept iteratively until the entanglement of the MPO is minimized, leaving two unitary disentangling circuits.}

\item{Compute a new MPO representing the remaining part of the original MPO that cannot be disentangled. This process is described by the equation
\beq
{\rm MPO}_{{\rm old}} \approx U \times {\rm MPO}_{{\rm new}} \times V^\dagger,
\eeq
where ${\rm MPO}_{{\rm old}}$ is the original MPO decomposition of the weight matrix $W$, ${\rm MPO}_{{\rm new}}$ is the ``remaining" MPO, and $U$, $V^\dagger$ are the unitary quantum circuits of disentanglers. Since ${\rm MPO}_{{\rm old}}$ is not necessarily Hermitian, we have that $U \neq V$ in general. Also, since $U$ and $V^\dagger$ remove entanglement from the original MPO, it follows that
\beq
\chi_{{\rm new}} \le \chi_{{\rm old}},
\eeq
where $\chi_{{\rm old}}$ and $\chi_{{\rm new}}$ are the bond dimensions of the old and new MPOs, respectively. The new MPO can be computed using the equation
\beq
{\rm MPO}_{{\rm new}} \approx U^{\dagger} \times {\rm MPO}_{{\rm old}} \times V.
\eeq
In practice, ${\rm MPO}_{\rm new}$ is here obtained by contracting the disentangling circuits $U^\dagger$ and $V$ with ${\rm MPO}_{\rm old}$ and truncating the bond dimension by singular value decomposition.}
\end{enumerate} 

This procedure is advantageous because it ensures a shallow quantum circuit combined with a low-dimensional MPO. This method can be applied uniformly across all deep layers of the LLM, allowing us to encode existing LLMs into this hybrid classical-quantum architecture.

\subsection{Enhancing the model beyond the classical baseline}
\label{sec:enhance}
Second, to enhance the original classical LLM we extend the quantum circuits for $U$ and $V^\dagger$ and increase the bond dimensions for ${\rm MPO}_{{\rm new}}$. The model parameters can be optimized through various techniques. Specifically, the unitaries in the quantum circuits $U$ and $V^\dagger$ can be optimized variationally using a self-consistent method similar to the Variational Quantum Eigensolver (VQE) algorithm~\cite{vqe}. Similarly, the tensors in the MPO can be retrained using distributed tensor network retraining techniques, as implemented in Ref.~\cite{compactifai}.

A complementary and particularly effective way to surpass the original model is to append an additional trainable unitary to the disentangled layer, which we refer to as a \emph{Cayley unitary adapter}~\cite{aizpurua2026quantumenhanced}: a block-diagonal orthogonal matrix $U_\theta$, parameterized through the Cayley transform, whose blocks act on consecutive groups of input features. The augmented layer then computes $W\,U_\theta(\mathbf{x})$, and since $U_\theta$ is initialized to the identity the hybrid model reproduces the original layer exactly before any training. Keeping every other parameter of the network frozen, only this adapter is trained, by supervised fine-tuning with knowledge distillation from the original model. Its single hyperparameter is the block size $b$, which sets the number of trainable parameters and hence the expressivity of the adapter; it may be placed on the input side of the layer, on the output side, or on both.

\subsection{Deployment on quantum hardware}
\label{sec:deployment}
Crucially, when deploying this architecture on actual quantum hardware, the input to the quantum circuits must be encoded as a quantum state, and their output is obtained through sampling via qubit measurements. Here we load the input state by amplitude encoding, although other schemes such as angle encoding or tensor network state preparation could also be tested. After executing the optimized quantum circuit for $U$, sampling is performed to reconstruct the input state as accurately as possible for ${\rm MPO}_{{\rm new}}$, which can also be encoded as a tensor network. Likewise, the input to $V^\dagger$ is encoded as a quantum state and the output is estimated via sampling. While these steps introduce additional approximations to the model, they can be controlled and compensated by increasing the depth of $U$ and $V^\dagger$ and the bond dimensions of ${\rm MPO}_{{\rm new}}$.

\section{Results}
\label{sec:results}
In all experiments below, we use a compressed version of SmolLM2~\cite{smollm2}, obtained with our tensor-network compression method from Ref.~\cite{compactifai}, and replace the self-attention weight matrix of the tenth layer, a linear layer of dimension $(576, 192)$ (arbitrary choice). This dimension sets the maximum gate-size of the $U$ and $V$ quantum circuits: embedding a dimension $d$ requires $\lceil \log_2 d \rceil$ qubits, i.e. the smallest power of two not smaller than $d$, which yields $(10, 8)$ qubits for $(576, 192)$, respectively.

Throughout this section we deliberately restrict the substitution to a single layer. This is a controlled proof of concept: it isolates the effect of the disentangling procedure and of the trained adapter on the full-model perplexity, which would otherwise be confounded by the simultaneous modification of many layers. Since the encoding of Sec.~\ref{sec:encoding} acts on one weight matrix at a time and is independent of the rest of the network, it can be applied layer by layer, and we expect the memory savings to accumulate across layers; a systematic study of the full-model substitution, where the interplay between modified layers and the associated retraining cost become relevant, is left for future work.


\subsection{Disentangling}
\label{sec:disentangling}
First, we analyze the disentangling process for a wide range of architectures for the $U$ and $V$ quantum circuits. We vary the size of the gates we implement in $U$, $k_u$: from 2 qubits to 10 qubits; the size of the gates we implement in $V$, $k_v$: from 2 qubits to 8 qubits; and the number of layers consisting of these gates that we implement at $U$ and $V$, $L$: from 1 to 500 for some cases.

\begin{figure}[tb]
    \centering
    \includegraphics[width=\linewidth]{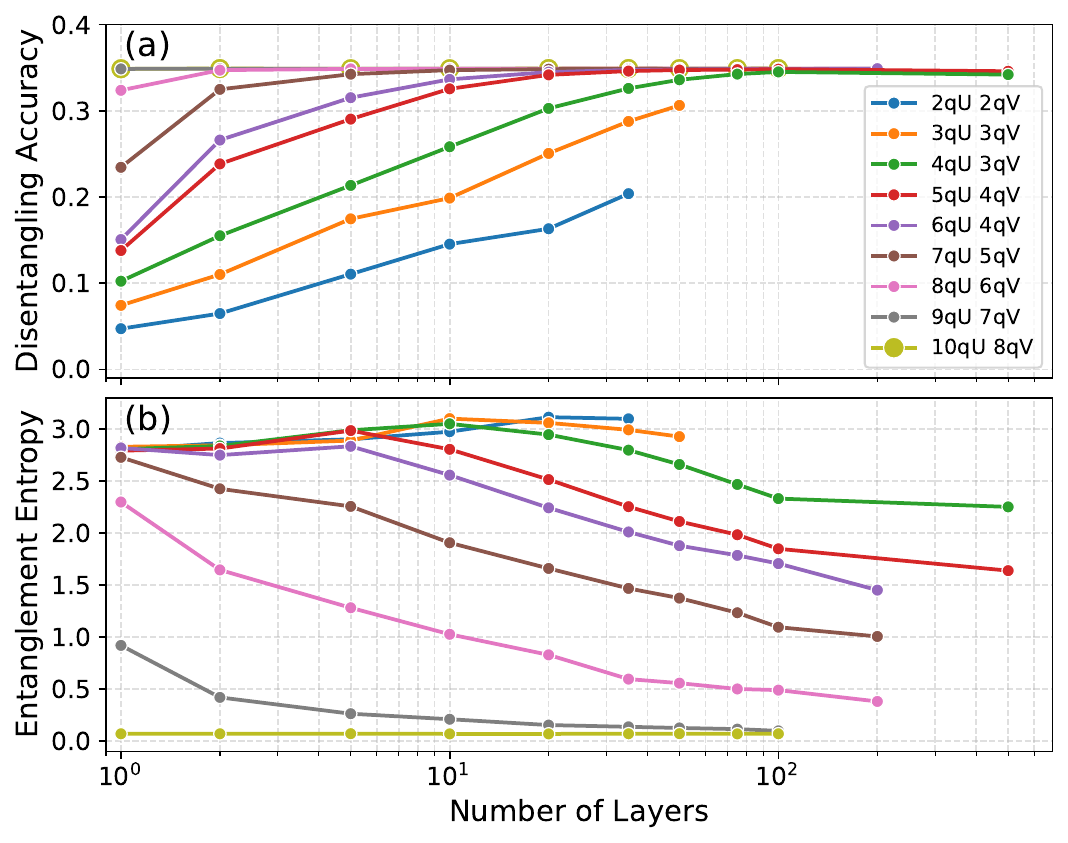}
    \caption{(a) Disentangling accuracy and (b) Entanglement entropy as a function of the number of layers, for nine gate-size combinations specified in the legend. Markers of case (10qU, 8qV) are enlarged in panel (a) for better visualization.}
    \label{fig:disentangling}
\end{figure}

In Fig.~\ref{fig:disentangling}, we represent the disentangling accuracy and the entanglement entropy for nine pairs of $U$ and $V$ gate sizes as a function of the number of layers $L$. The disentangling accuracy measures how close the layer is to being fully disentangled, i.e. to a product (bond-dimension-$1$) operator. We define the target ${\rm MPO}_{\rm target}$ as ${\rm MPO}_{\rm old}$ truncated to bond dimension $1$, and take the accuracy to be the normalized overlap between ${\rm MPO}_{\rm old}$ with the optimized disentanglers applied, $U^\dagger\,{\rm MPO}_{\rm old}\,V$, and this target,
\beq
\mathcal{A} = \frac{{\rm Tr}\!\left[{\rm MPO}_{\rm target}^{\dagger}\,U^{\dagger}\,{\rm MPO}_{\rm old}\,V\right]}{\|{\rm MPO}_{\rm old}\|\;\|{\rm MPO}_{\rm target}\|}\,,
\eeq
which lies in $[0,1]$ and equals $1$ when ${\rm MPO}_{\rm old}$ disentangles exactly into the product form. The disentanglers $U$ and $V^\dagger$ are precisely those that maximize $\mathcal{A}$ (see Appendix~\ref{app:disentangling}). The entanglement entropy is the von Neumann entropy across the bonds of the disentangled MPO: denoting by $\{s_\alpha\}$ the Schmidt values across a given bond and $p_\alpha = s_\alpha^2/\sum_\beta s_\beta^2$ the corresponding normalized weights, the entropy of that bond is $S = -\sum_\alpha p_\alpha \ln p_\alpha$, and we report its mean over all bonds.

As shown in Fig.~\ref{fig:disentangling}, the accuracy increases (saturating at large $L$) while the entropy decreases monotonically with the number of layers $L$, the two being strongly anticorrelated: a less entangled MPO is, by construction, better approximated by the bond-dimension-$1$ target. How quickly this happens is governed by the gate size. For the largest gates the layer is essentially disentangled already at $L=1$; the $(10q_U,8q_V)$ case reaches $\mathcal{A}\approx0.35$ with a residual entropy of only $\sim0.07$, and the next few sizes converge within a handful of layers. The smallest gates, by contrast, barely reduce the entanglement at all: the $(2q_U,2q_V)$ case stays close to its initial entropy ($\sim2.9$) and only reaches $\mathcal{A}\approx0.20$ even at $L=35$. In all cases the dependence on $L$ is approximately logarithmic, so the gains become marginal as layers are added, and the accuracy saturates well below unity (around $0.35$ at bond dimension $1$).

This metric is pessimistic, however: it is an overlap with the bond-dimension-$1$ truncation of the \emph{original} weight matrix, and as we show below (Table~\ref{tab:memory}) truncating the disentangled operator itself to low bond dimension barely affects the perplexity. The large-gate circuits indeed encapsulate the layer into a nearly product (bond-dimension-$1$) MPO flanked by two quantum circuits, confirming that the procedure is effective. The gate size is thus a trade-off: small gates are cheap per layer but disentangle slowly and require long optimizations, whereas large gates disentangle in very few layers but yield circuits whose depth grows rapidly once transpiled to hardware-native gates. We stress that these results are not pushed to convergence: deeper circuits (more layers) and a tighter optimization tolerance would drive the entanglement entropy still lower, but we restricted the experiments owing to limited compute time and hardware availability.

\subsection{Bond dimension}
\label{sec:bond-dimension}
\begin{figure}[tb]
    \centering
    \includegraphics[width=\linewidth]{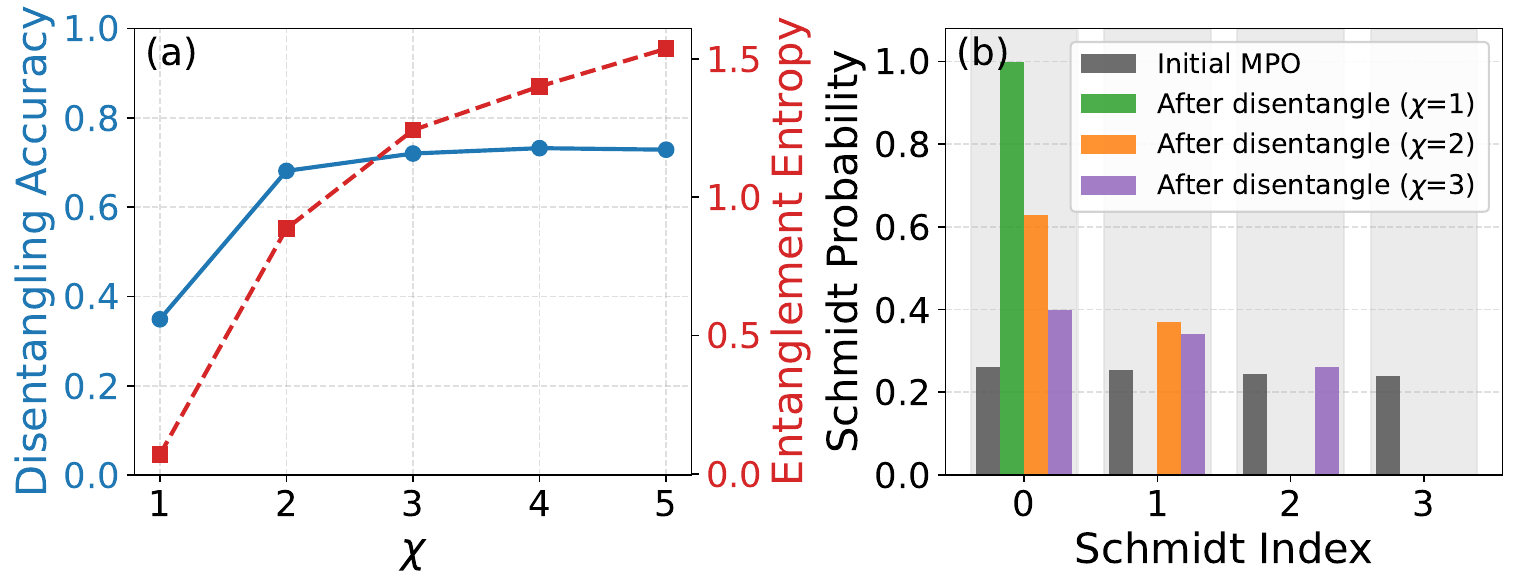}
    \caption{(a) Disentangling accuracy (blue) and Entanglement entropy (red) as a function of the bond dimension of the target MPO. (b) Histogram of the Schmidt probability at bond $7$ for the original MPO (dark grey) and after the disentangling process with bond dimensions: 
    $\chi = 1$ (green), $\chi = 2$ (orange) and $\chi = 3$ (purple). The replaced layer of the model considered for both panels is characterised by one 10-qubit gate as U circuit and one 8-qubit gate as V circuit, this is: 10qU, 8qV and 1 layer.}
    \label{fig:bond_dims}
\end{figure}

The accuracy ceiling found above is set by the bond dimension of the target. To show this, we fix the most expressive architecture, $(10q_U,8q_V)$ with $L=1$, which is already disentangled at maximal accuracy, and increase the bond dimension $\chi$ of ${\rm MPO}_{\rm target}$ [Fig.~\ref{fig:bond_dims}(a)]. The accuracy rises steeply from $\mathcal{A}\approx0.35$ at $\chi=1$ to $0.68$ at $\chi=2$ and $0.72$ at $\chi=3$, and then plateaus around $0.73$ for $\chi\geq4$. The remaining gap to unity is a consequence of the intrinsic complexity of the self-attention weight matrix, which is not captured exactly even by a low-bond-dimension MPO. The entanglement entropy behaves in the opposite way, growing monotonically with $\chi$ (from $0.07$ at $\chi=1$ to $1.54$ at $\chi=5$). This is expected: at $\chi=1$ the disentangler is forced to concentrate the weight of every bond into a single Schmidt value, driving the entropy to nearly zero, whereas allowing $\chi>1$ preserves several Schmidt values and hence retains larger entanglement. Figure~\ref{fig:bond_dims}(b) makes this explicit at bond $7$, whose four nearly degenerate Schmidt weights in the original MPO, $\{0.26,0.255,0.245,0.24\}$, collapse onto a single index ($p_0=1$) after disentangling at $\chi=1$, split into two ($0.63,0.37$) at $\chi=2$, and into three ($0.40,0.34,0.26$) at $\chi=3$.

\subsection{Classical memory}
\label{sec:classical-memory}
Disentangling concentrates the layer's correlations into the circuits $U$ and $V^\dagger$, leaving a residual operator ${\rm MPO}_{\rm new}$ that is highly compressible. Table~\ref{tab:memory} reports the full-model perplexity (PPL) as ${\rm MPO}_{\rm new}$ is truncated to bond dimension $\chi$ for the $(10q_U,8q_V,L{=}1)$ layer. As anticipated in Sec.~\ref{sec:disentangling}, the low disentangling accuracy at $\chi=1$ ($\mathcal{A}\approx0.35$) is a pessimistic figure of merit and does not translate into a perplexity loss: collapsing ${\rm MPO}_{\rm new}$ all the way to a product MPO ($\chi=1$, just $36$ classical parameters) raises the PPL by only $0.26\%$, and $\chi=2$ ($132$ parameters) by $0.15\%$. The classical tensor-network footprint of the layer thus shrinks from the $110{,}592$ parameters of $W$ to a few tens of parameters, more than three orders of magnitude smaller, with the perplexity preserved to within $0.3\%$. The removed entanglement is carried by $U$ and $V^\dagger$; this classical-memory saving is realized when these circuits are executed on quantum hardware rather than stored as dense tensors. The disentanglers are thus not a hidden classical cost but the quantum resource itself: as circuits they carry at least as many trainable parameters as $W$, realized on the QPU as the model's added expressive capacity rather than stored as classical weights. For comparison, truncating the same layer's undisentangled ${\rm MPO}_{\rm old}$ to bond dimension one degrades the perplexity by $9.7\%$, more than thirty times the $0.26\%$ cost of the disentangled truncation.
Thus, the disentangling circuits, not the MPO truncation, are what enable the compression.

\begin{table}[t]
\caption{Word-level WikiText perplexity (PPL) of the full model as the disentangled operator ${\rm MPO}_{\rm new}$ of the $(10q_U,8q_V,L{=}1)$ layer is truncated to bond dimension $\chi$. $\Delta$PPL is relative to the original model (weight matrix $W$: $110{,}592$ parameters, PPL $35.292$).}
\label{tab:memory}
\begin{ruledtabular}
\begin{tabular}{cccc}
$\chi$ & ${\rm MPO}_{\rm new}$ params & PPL & $\Delta$PPL \\
\hline
$1$ & $36$ & $35.383$ & $+0.26\%$ \\
$2$ & $132$ & $35.346$ & $+0.15\%$ \\
$5$ & $696$ & $35.307$ & $+0.04\%$ \\
$10$ & $2{,}356$ & $35.314$ & $+0.06\%$ \\
$50$ & $36{,}948$ & $35.300$ & $+0.02\%$ \\
exact & $401{,}956$ & $35.292$ & $0.00\%$ \\
\hline
$W$ & $110{,}592$ & $35.292$ & --- \\
\end{tabular}
\end{ruledtabular}
\end{table}

\subsection{Perplexity without the adapter}
\label{sec:ppl-no-adapter}
Having established that the layer can be disentangled, we next verify that this rewriting leaves the quality of the model intact. We evaluate the perplexity (PPL) of the full model before and after disentangling, for all nine gate sizes and for $L=1$ and $L=2$. In every case the PPL is essentially unchanged: the baseline value, $35.29$, is recovered after substituting and disentangling the layer to within a relative difference of order $10^{-3}\%$ (at most $0.003\%$). Disentangling by itself therefore neither improves nor degrades the model, consistent with the layer being faithfully re-expressed as two quantum circuits and an MPO. Any change in PPL must then originate from training the added unitary adapter, which we examine next.

\subsection{Training the unitary adapter}
\label{sec:adapter}
So far the quantum circuits have only re-expressed the existing layer. To surpass the original model we now train the Cayley unitary adapter introduced in Sec.~\ref{sec:enhance}. We place a single adapter on the input side of the disentangled layer and sweep its block size from $b=2$ (one-qubit blocks, $288$ trainable parameters) up to $b=576$, a single block spanning the full input dimension of the layer, keeping every other parameter of the network frozen.

\begin{figure}[tb]
    \centering
    \includegraphics[width=\linewidth]{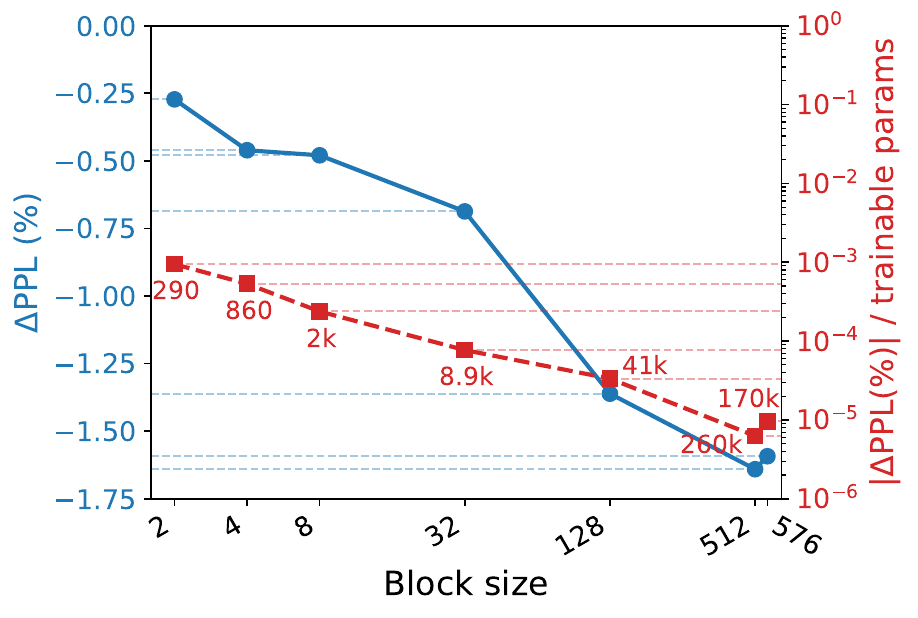}
    \caption{In blue, PPL relative difference ($\Delta$PPL) between the post-trained model and the baseline model (without replacing the weight matrix by our architecture); and in red, absolute value of $\Delta$PPL divided by the number of trainable parameters, both as a function of the block size of the trained unitaries. The number of trainable parameters for each block size is noted in red close to each point. The replaced layer of the model considered is characterised by one 10-qubit gate as U circuit and one 8-qubit gate as V circuit, this is: 10qU, 8qV and 1 layer.}
    \label{fig:preunitaries}
\end{figure}

Training the adapter lowers the perplexity well below the baseline, and the improvement grows with the block size [Fig.~\ref{fig:preunitaries}, blue]. For the representative $(10q_U,8q_V,L=1)$ configuration shown, the relative PPL difference reaches about $-1.6\%$ for the largest blocks; here, $b=512$ is the largest power-of-two block compatible with a $9$-qubit encoding, while $b=576$ corresponds to a single unconstrained block spanning the full hidden dimension. This improvement is more than three orders of magnitude larger than the $\sim10^{-3}\%$ shift produced by disentangling alone. The four other gate-size configurations we tested ($2$-$2$, $3$-$3$, $6$-$4$ and $9$-$7$) follow essentially the same curve, so the gain is set by the adapter rather than by the disentangler architecture. The largest blocks yield the biggest absolute improvement but at a steep parameter cost; normalizing $|\Delta{\rm PPL}|$ by the number of trainable parameters [Fig.~\ref{fig:preunitaries}, red] shows that the smallest block, $b=2$, is by far the most parameter-efficient, delivering a $\sim0.25\%$ improvement with only $288$ parameters, of order $10^{-3}\%$ per parameter.

\subsection{Quantum inference}
\label{sec:quantum-inference}
As a proof of concept, we run the hybrid model end-to-end with part of its circuits executed on real quantum hardware. Taking a trained model (disentanglers $U$, $V^\dagger$ and a trained unitary adapter), we perform inference on the fixed prompt ``What is 9 times 6?'', routing a chosen subset of the disentangler gates to a quantum backend while the rest of the network runs classically. The model whose disentanglers are built from two-qubit gates ($k_u=k_v=2$) returns the correct answer when these gates are executed on a superconducting quantum processor (IBM Heron), whereas configurations with larger gates cannot be run on hardware, as our QPU pipeline currently supports only two-qubit gates. This confirms that the disentangled architecture is compatible with present-day quantum devices; a systematic hardware study of the trained unitary adapters, including the role of noise, is reported separately in Ref.~\cite{aizpurua2026quantumenhanced}.

\section{Discussion and conclusions}
\label{sec:discussion}
We have introduced a method to encode a layer of a classical LLM into a hybrid quantum-classical architecture: a tensor-network disentangling procedure reproduces the layer while moving part of its parameters into two variational quantum circuits and leaving a low-bond-dimension MPO. Because the construction reproduces the original layer, the hybrid model departs from the classical baseline rather than from a randomly initialized ansatz. On a compressed SmolLM2 model, the disentangling step preserves the model's perplexity (PPL) to within a relative difference of order $10^{-3}\%$, while allowing the residual MPO to be truncated all the way to bond dimension one: this reduces the classical parameter count of the layer from $110{,}592$ to $\sim36$, more than three orders of magnitude, at a perplexity cost below $0.3\%$. Training the added unitary adapter then lowers the PPL by up to $\sim$1.6\% below the baseline, and we demonstrate hybrid inference of the resulting model on a real quantum processor as a proof of concept.

Notably, our method can also be combined with standard compression techniques for classical LLMs such as quantization~\cite{Quantization}, distillation~\cite{Distilling}, pruning~\cite{Pruning}, and matrix factorizations such as low-rank approximations~\cite{Lora}. Additionally, these results indicate that quantum circuits with tens of qubits and hundreds of layers should suffice to improve some of the current open-source LLMs. This is particularly promising, since it matches also the roadmap of quantum hardware providers such as IBM~\cite{ibm_quantum_roadmap_2024}.


We emphasize that the aim of this paper is not to demonstrate a quantum advantage in inference cost or speed, but to initiate concrete algorithms for this hybrid classical-quantum framework, one that could eventually reach scaling regimes beyond purely classical tensor-network representations. Our disentangling procedure replaces classical parameters of a pretrained model with parameters encoded in a quantum circuit, and we therefore study expressivity as a distinct scaling resource enabled by the intrinsically quantum circuit ansatz. Our vision is that Quantum LLMs, like the ones described in this paper, may become in the mid-term one of the first practical applications of noisy quantum computers.

\begin{acknowledgments}
We acknowledge Donostia International Physics Center (DIPC), Ikerbasque, Basque Government, Diputaci\'on de Gipuzkoa, European Innovation Council (EIC) and Spanish Government for constant support, as well as insightful discussions with the team from Multiverse Computing S. L.
\end{acknowledgments}


\bibliography{references} 

\appendix

\section{Explicit disentangling algorithm}
\label{app:disentangling}

This appendix summarizes the explicit variational algorithm used to compute the disentangling circuits, introduced in Ref.~\cite{fmps-tjwy} for classical neural networks and adapted here to the MPOs that arise from the weight layers of an LLM.

The construction replaces the weight matrix $W$ by a MPO, ${\rm MPO}_{\rm old}$, that is then disentangled into
\[
{\rm MPO}_{\rm old} \approx U \, {\rm MPO}_{\rm new} \, V^\dagger, \qquad \chi_{\rm new} < \chi_{\rm old},
\]
where $U, V^\dagger$ are the disentangling quantum circuits run on hardware and ${\rm MPO}_{\rm new}$ is a compact MPO that stays in the classical network. The two steps are detailed below.

The first step tensorises the weight matrix $W$ into ${\rm MPO}_{\rm old}$ following the tensor-network compression of neural-network layers~\cite{tnn0, tnn3} and of LLMs~\cite{compactifai}. This is not a plain low-rank approximation: pretrained weights are typically close to full rank, so a naive truncation degrades accuracy. The model is instead \emph{healed} (briefly retrained) after the substitution, yielding an ${\rm MPO}_{\rm old}$ that reproduces $W$ to within the model's performance. With ${\rm MPO}_{\rm old}$ fixed, the remaining task is to disentangle it, which first requires fixing a circuit ansatz.

Ideally we fully disentangle the MPO ($\chi_{\rm new}=1$), which corresponds to a disentangling channel $\Phi$ acting as $\Phi({\rm MPO}_{\rm old}\,{\rm MPO}_{\rm old}^\dagger)=\bigotimes_i P_i$, a product of positive matrices. Its Kraus decomposition
\begin{equation}\label{eq:kraus}
\Phi(\cdot) = \sum_k U_k (\cdot) U_k^\dagger,
\end{equation}
yields a set of disentangling unitaries $\{U_k\}$, each approximated by a quantum circuit. We keep a single circuit ($k=1$), identifying the left disentangler $U\equiv U_1$, and allow an independent right circuit $V^\dagger\neq U^\dagger$, since ${\rm MPO}_{\rm old}$ is in general non-Hermitian. For the ansatz we use a brickwall of few-body gates, optionally interleaved with single-qubit layers, and restrict to real orthogonal gates for numerical stability; complex gates would be more expressive and are left to future work.

To keep the transpiled physical circuits shallow, some or all two-qubit gates can be fixed to hardware-native CNOTs, optimizing only the remaining single-qubit gates. In the extreme case the two-qubit gates form a frozen CNOT brickwall and only single-qubit gates are trained, which still provides substantial disentangling at a much lower transpilation cost~\cite{fmps-tjwy}.

\begin{figure}[t]
    \centering
    \includegraphics[width=0.95\linewidth]{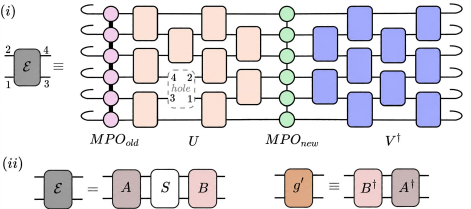}
    \caption{Variational optimization of disentanglers. (i) The \emph{environment tensor} $\mathcal{E}_g$, Eq.~\ref{eq:env}, corresponding to a gate $g$ being optimized is obtained by removing that tensor from the circuit, leaving a hole with open indices, and contracting the remaining tensors that appear in Eq.~\ref{eq:env}. (ii) The singular value decomposition of $\mathcal{E}_g$ is performed, \(\mathcal{E}_g = A S B\). The updated gate is obtained as $g' = B^\dagger A^\dagger$.}
    \label{fig:disentangler_optimization}
\end{figure}

The gates are found by maximizing the normalized overlap
\begin{equation}\label{eq:var_loss}
\mathrm{Tr}\left({\rm MPO}_{\rm old} (U \, {\rm MPO}_{\rm new} \, V^\dagger)\right) \big/ \left(\|{\rm MPO}_{\rm old}\| \, \|{\rm MPO}_{\rm new}\|\right),
\end{equation}
as in the determination of MERA disentanglers~\cite{MERA, disentanglers}. In the experiments reported in the main text we specialize to a fully disentangled target, fixing ${\rm MPO}_{\rm new}$ to the bond-dimension-$1$ truncation of ${\rm MPO}_{\rm old}$ (denoted ${\rm MPO}_{\rm target}$ there) and optimizing only the circuits $U$ and $V^\dagger$. Starting from random gates, we sweep over both circuits; for each gate $g$ (say, in $U$) we compute its environment tensor
\begin{equation}\label{eq:env}
\begin{split}
\mathcal{E}_g \equiv \mathrm{Tr}\left({\rm MPO}_{\rm old} \big(U^{\,g \rightarrow \mathrm{hole}} {\rm MPO}_{\rm new} V^\dagger\big)\right) \\
\big/ \left(\|{\rm MPO}_{\rm old}\| \, \|{\rm MPO}_{\rm new}\|\right),
\end{split}
\end{equation}
i.e. the network with $g$ replaced by an open hole, contracted with the rest (Fig.~\ref{fig:disentangler_optimization}). The environment is the local gradient of the overlap with respect to $g$, so each update is a gradient step that exploits the local structure and, in practice, converges faster and more robustly than full-network gradient descent~\cite{fmps-tjwy}. The gate is then updated from the singular value decomposition $\mathcal{E}_g = A S B$ as
\begin{equation}\label{eq:update}
g' \equiv B^\dagger A^\dagger.
\end{equation}

The cost is dominated by contracting the circuit-MPO network for each environment, which grows with circuit depth. It is exact for the shallow circuits used here and, for deeper circuits, can be approximated with boundary-MPS or renormalization-group schemes~\cite{tn1, tn2}. We disentangle the compressed MPO rather than the dense $W$ directly (for instance through the unitary factors of its QR or singular-value decomposition), because the small-bond-dimension MPO is far less entangling and yields shallower circuits.

Alternative approaches to introduce quantum circuits, such as using the polar decomposition $W = U \times P$ of the weight matrix, where $U$ is unitary and $P$ is positive-definite, are also possible. However, our numerical tests indicate that this approach results in a non-shallow quantum circuit for $U$ and a $P$ with a very large bond dimension, due to the positivity constraint. In this context, the disentangler approach offers the most compact possible representation of $W$.

\end{document}